\documentclass[twocolumn,secnumarabic,amssymb, nobibnotes, aps, prl]{revtex4-2}

\usepackage{amsmath}
\usepackage{graphicx}
\usepackage[colorlinks=true, allcolors=blue]{hyperref}

\begin{document}

\title{Yield Stress Fluids Solidifying in Capillary Imbibition}%

\author{Hanul Kim, Siyoung Q. Choi}%
\email[]{sqchoi@kaist.ac.kr}
\affiliation{Department of Chemical and Biomolecular Engineering, Korea Advanced Institute of Science and Technology (KAIST), 291 Daehak-ro, Yuseong-gu, Daejeon 34141, Republic of Korea}
\date{August 2023}%

\begin{abstract}
When subjected to an external stress that exceeds the yield stress ($\sigma_\mathrm{Y}$), yield stress fluids (YSFs) undergo a solid-to-liquid transition. Despite the extensive studies, there has been limited attention to the process of liquid-to-solid transition. This work examines the solidification of YSFs through capillary imbibition, easily observed in the processes of wetting, coating, spreading, and wicking. During gradual deceleration of the capillary rise, YSFs display an unexpected flowing behavior, even when subjected to stresses below the $\sigma_\mathrm{Y}$. We propose a model with numerical solutions based on rheological properties of YSFs and slip to capture this unusual, yet universal behavior.
\end{abstract}
\maketitle

Yield stress fluid (YSF) keeps its shape like a solid until a sufficiently large stress ($>\sigma_\mathrm{Y}$) initiates a liquid-like flow \cite{Bonn2017a, Coussot2002}. YSFs are widely found in biofluids (cell medium and mucus) \cite{Hu2020, hu2015}, as well as in various industries such as paintings, cosmetics, and foods \cite{Sun2009, Malkin2017}. For YSFs, the solid-to-liquid change is the key feature commonly denoted as yielding \cite{Moller2009}. The yielding depends on the strain rate and stress, and shows a gradual transition rather than a sudden event \cite{Oldroyd1947,Coussot2021a,Donley2020, Kamani2021,Bonn2009,coussot2018}. Since plastic deformation takes place after elastic response, stress overshoot appears as a consequence of microstructural rearrangement and transient/permanent shear banding follows due to its inhomogeneity \cite{Benzi2021a,serial2021,coussot1995,Goyon2008}.

In contrast to yielding, solidification of YSF, a liquid-to-solid transition where the stress decreases below $\sigma_\mathrm{Y}$, is rarely studied although it is found everywhere in capillarity-driven flows such as wetting, coating, spreading, and wicking \cite{DeGennes2004,Martouzet2021, Jorgensen2017}. The mismatch between dynamic and static yield stresses suggests that the solidifying transition is expected to differ from the yielding transition \cite{Cuny2021}. However, the solidification process of YSFs still remains unclear.

\begin{figure}[b]
\includegraphics[width=1\linewidth]{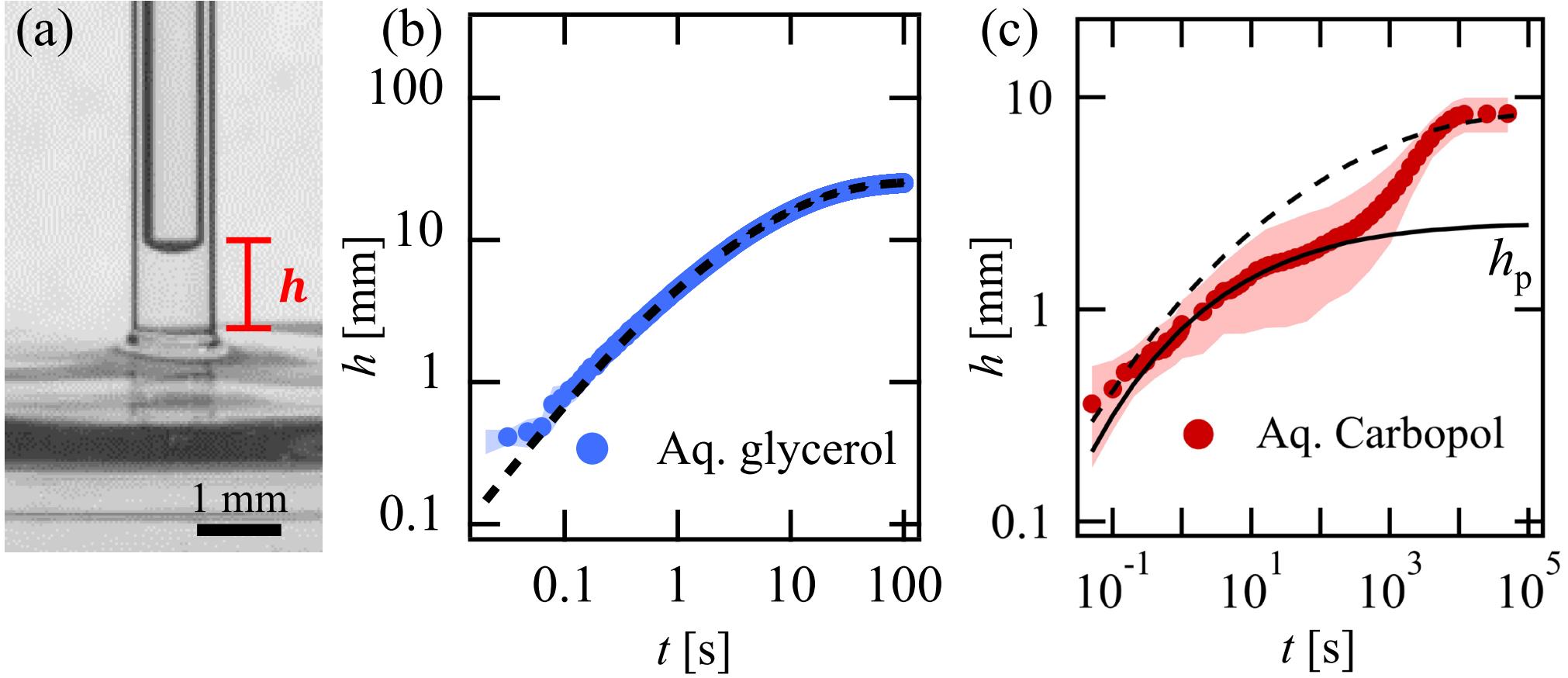}
\caption{\label{fig:Fig1}(a) Schematic for capillary rise experiment. Capillary rise curves of (b) 85 $\%$ glycerol (Newtonian fluid, $\sigma_\mathrm{Y}=0$) and (c) 0.12 wt. $\%$ Carbopol (YSF, $\sigma_\mathrm{Y}=10$ Pa). Equation (1) is plotted as a dashed line. The solid line is equation (1) connected with Herschel–Bulkley (HB) model  (equation (2)). The height of plateau is denoted with $h_\mathrm{p}$. The shade is error bar.}
\end{figure}

In this letter, our primary focus is the solidification of YSFs in capillary rise (Fig. \ref{fig:Fig1}(a)). From the conservation of momentums acting on the liquid (density $\rho$), the wall shear stress on capillary inner surface with radius $R$ is function of rise height ($h(t)$) as expressed by 
\begin{equation}
\sigma_\mathrm{w}\left(t\right)=\frac{\mathit{\Gamma}\cos \theta_\mathrm{E}}{h(t)}-\frac{\rho gR}{2},
\end{equation}
where $\mathit{\Gamma}$ is the surface tension of liquid-air interface; $\theta_\mathrm{E}$ is the equilibrium contact angle between the liquid and the capillary; $g$ is the gravitational constant \cite{Lucas1918, Washburn1921,Rideal1922}. Because $h(t)$ increases by time, $\sigma_{w}\left(t\right)$ finally converges to zero for typical constitutive equations (Newtonian or shear thinning fluid). This corresponds to a monotonic rise curve of aqueous glycerol in Fig. \ref{fig:Fig1}(b). YSFs, however, undergo solidification at $h_\mathrm{p}=\mathit{\Gamma}\cos \theta_\mathrm{E}/\left(\sigma_\mathrm{Y}+\rho gR/2\right)$, the height where $\sigma_\mathrm{w}=\sigma_\mathrm{Y}$. This results in a novel flow pattern of traversing a plateau prior to reaching a final height as Fig. \ref{fig:Fig1}(c). This contradicts a previous study that says YSF stops at $h_p$ \cite{Geraud2014}. Instead, we found that solidification slows down YSFs at first, but is followed by another onset of capillary rise, which is dominated by slip. Our proposed model predicts the complete capillary rise of YSFs quite accurately, and explains what determines the solid-to-liquid reentrance based on the relevant length and time scales.

Capillary rise experiments are performed using a glass capillary with radius 0.85 or 0.42 mm. For YSFs, Carbopol and Xanthan aqueous solutions are used. The solution in a Petri dish itself raised slowly until it touches the capillary's bottom to initiate imbibition. The entire capillary rise is recorded with a CCD camera and analyzed with a custom MATLAB code \cite{Kim2020}. See SI for detailed methods \cite{SI}.

Capillary rise curves of Newtonian (glycerol) and yield stress fluids (Carbopol) are compared in Figs. \ref{fig:Fig1}(b) and (c). Newtonian fluids exhibit monotonic increase until it approaches its ultimate equilibrium height ($h_\mathrm{E}=2\mathit{\Gamma} \cos \theta_\mathrm{E}/\rho gR$), wherein the Laplace pressure is balanced by gravity. When Hagen-Poiseuille flow is assumed \cite{Sutera1993}, the model curve accords with Newtonian fluid's data. However, with a simple power law fluid (i.e., shear thinning viscosity), the model is unable to match for YSFs. The most likely reason for this gap could be the absence of yield stress ($\sigma_\mathrm{Y}=0$) and inability to account for the impact of yielding or solidification.

Accordingly, a well-known Herschel–Bulkley (HB) model is implemented to equation (1). For $\sigma>\sigma_\mathrm{Y}, \sigma=\sigma_\mathrm{Y}+K\dot{\gamma}^\mathrm{n}$, where $\dot{\gamma}$ is the shear rate; $K$ is the consistency parameter; $n$ is the thinning power. When $\sigma\le\sigma_\mathrm{Y}, \dot{\gamma}=0$. The apparent rise velocity $\dot{h(t)}$ is obtained by averaging the shear rate in radial direction; $\dot{h}(t)=\int_{0}^{R}2\pi r\dot{\gamma}(r,t)\,dr$ \cite{Sochi2008, Perez-Gonzalez2012}. By replacing the shear rate with HB model, it rearranges and explicitly links $\dot{h}(t)$ to the wall shear stress ($\sigma_{\mathrm{w}}(t)$) as
\begin{equation}
\dot{h}\left(t\right)=R\left(\frac{\sigma_\mathrm{Y}}{K}\right)^{s-1}f\left(\epsilon\left(t\right),s\right), 	
\end{equation}
where $f=\epsilon(t)\left(\frac{1-\epsilon(t)}{\epsilon(t)}\right)^s\left\{\frac{\left(1-\epsilon(t)\right)^2}{s+2}+\frac{2\epsilon(t)(1-\epsilon(t))}{s+1}+\frac{\epsilon(t)^2}{s}\right\}$; $\epsilon\left(t\right)=\frac{\sigma_\mathrm{Y}}{\sigma_\mathrm{w} \left(t\right)}<1$; $s=1/n+1$. For regime $\epsilon(t)\ge1, \dot{h}=0$. Equation (2) can be reduced to shear thinning model by putting $\sigma_\mathrm{Y}=0$. For the simple shear thinning fluid, the stress will decrease monotonically, but for YSF, flow will cease at $h_\mathrm{p}$.

With the addition of equation (2), the numerical solution is shown as a solid line in Fig. \ref{fig:Fig1}(c). The plateau height, denoted as $h_\mathrm{p}$, was precisely determined by the model. This indicates that the plateau is a result of the solidification process of YSF.

Interestingly, stress overshoot has negligible impact on the macroscale dynamics of capillary rise. Recent experimental and theoretical studies have demonstrated that the transition from solid to liquid is accompanied by a stress overshoot, as an evidence of plastic deformation of YSF \cite{Donley2020, Kamani2021}. In a case of the capillary rise, the good fit of equation (2) to the initial rising speed indicates that stress overshoot has little effect on the macroscale flow dynamics; With the sufficiently large Laplace stress ($\sigma \sim$ 100 Pa) at the beginning, the characteristic time scale for yielding rate ($\sim$ $10^{-2}$ s) is an order of magnitude faster than the time scale of our measurements ( $\ge$ $10^{-1}$ s) \cite{Benzi2021a}.

For each model curve, dynamic contact angle (DCA) theory is employed to account for the wetting friction of meniscus\cite{Kim2020,Primkulov2020}. The DCA description and its fitting are available in \cite{SI}. The YSF’s surface tension is estimated from the Laplace pressure at $h_\mathrm{E}$ due to the lack of reproducibility of the previously proposed method, which heavily relies on the selection of geometry \cite{Jorgensen2015}.

Despite the good agreement of HB model for the early time of Fig. \ref{fig:Fig1}(c), $h(t)$ after $h_\mathrm{p}$ is still unaddressed. The experimental curve exhibits a significant deviation from the saturation predicted by equation (2), and continues to rise until $h_\mathrm{E}$. The origin of this unexpected flow must be linked to what happens at the end of the L-S transition. Given that shear stress approaches to zero after $\sigma_\mathrm{Y}$, the solidification is distinct from the re-entrance effect under long-term creep test \cite{Landrum2016} and the thixotropic aging under the rest after shearing \cite{Coussot2014,choi2020,choi2021}.

One possibility is that YSFs may not undergo complete solidification. This idea comes from the analogy of delayed yielding and can be referred to as ‘delayed solidification’ \cite{Landrum2016}. While the classical HB model defines the state of matter solely based on shear stress, this approach emphasizes the rheological inhomogeneity of YSF. This approach incorporates local mobility as a function of shear stress and shear rate \cite{Benzi2021a,Benzi2021}. Even if there has been no theoretical discussion on solidification per se, the studies from the S-L transition can be mirrored to L-S. Since fluidization occurs even below the yield stress, solidification can also be delayed, making YSF a mixture of solid and liquid. Even if the yielding is quick at the beginning of the imbibition process, the solidification can be substantially slower since it takes place in distinct shear rate ($\dot{\gamma}$), stress ($\sigma$), and stress rate ($\dot{\sigma}$). Moreover, because the mobility diffuses out rapidly in fast yielding, its reverse process, solidification, might be slow.

The other possibility is YSF flows after $h_\mathrm{P}$ as a “gliding solid block”. If we assume a no-slip condition at the wall, Oldroyd-Prager formulation expects the negative rise velocity ($\dot{h}<0$) due to the elastic deformation below the yield stress \cite{Coussot2021a, Oldroyd1947,Hohenemser1932, Dinkgreve2017}, which is not observed in our experiments. However, the experiments can be explained if we consider the presence of a shallow lubrication layer near the wall, where high shear stress is concentrated. The slip of YSFs has been consistently postulated, irrespective of their yielding mechanism, chemical composition, geometry, or wall surface roughness \cite{Freydier2017,Shamu2020,Cox1986,Liu2018,Jalaal2015,Graziano2021,Pemeja2019,Zhang2017,Piau2007}. The slip has known to occur by the electrical repulsion between the dispersed component (e.g., polymers in our system) and the wall surface, so this layer is mostly the solvent \cite{Zhang2017}.

\begin{figure}[t]
\includegraphics[width=1\linewidth]{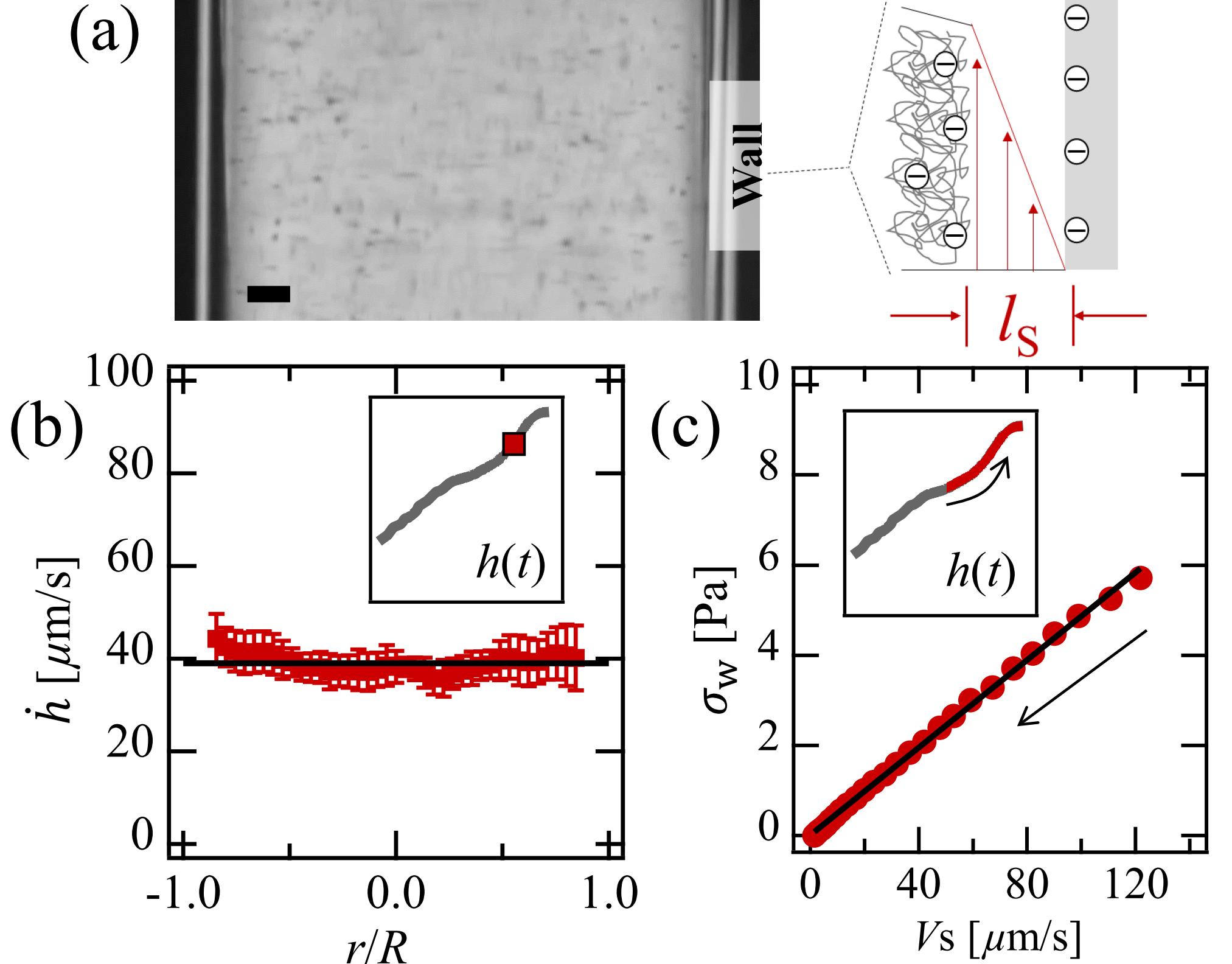}
\caption{\label{fig:Fig2}Slip demonstrations. (a) Image of 0.12 wt. $\%$ Carbopol inside glass capillary. Solution is seeded with 10 $\mu$m hollow glass spheres. Scheme on the right describes the depletion layer at the YSF-glass interface. (b) Particle image velocimetry (PIV) analysis result. Solid line is plug flow velocity profile from the bulk imbibition speed. (c) $\sigma_\mathrm{w}-V_\mathrm{s}$ plot of Xanthan 1.0 wt. $\%$. Solid line is the Newtonian slip model. Insets are $h(t)$ curves and colored markers denote corresponding regimes for (b) and (c). Scale bar is 100 $\mu$m.}
\end{figure}

In order to verify the aforementioned hypotheses, the flow velocity is visualized with the particle image velocimetry (PIV) method. We expect a plug flow in case of entire solid, otherwise a curved velocity profile for mixed solid-fluid state. To do experiments, hollow glass spheres are added to YSFs and observed under the optical microscope. The exemplary optical image is in Fig. \ref{fig:Fig2}(a). The velocity profile was analyzed during horizontal imbibition due to its convenience of imaging. This method remains valid as YSF solidifies even in the absence of gravity, and the penetration curve exhibits a plateau similar to that of the vertical rise \cite{SI}. The velocity profile depicted in Fig. \ref{fig:Fig2}(b) is indeed a plug, suggesting that the fluid undergoes complete solidification instead of existing in a state of mixture between solid and liquid.

To confirm the slip effect, the wall shear stresses are examined from the all the points after the plateau. Because the wall shear stress can be computed simply by the rise ($h$) using equation (1), independent of the constitutive relationship to the fluid viscosity. In Figure \ref{fig:Fig2}(c), it shows the wall shear stress linearly increases with the rise velocity, a behavior of Newtonian fluid, and the same relation has have been found in other YSFs as well  \cite{Zhang2017,SI}. In other word, even if the apparent viscosity changes, YSFs shares the similar \textit{slip} behavior at boundaries.

To implement \textit{slip}, Newtonian slip model \cite{Zhang2017} is added to equation (2). The slip layer has thickness $l_\mathrm{s}$, in which polymers are depleted. When its viscosity is assumed as the solvent viscosity ($\mu_\mathrm{solvent}$), the slip velocity can be written with wall shear stress proportionally as 
\begin{equation}
V_\mathrm{s}(t)=\frac{l_\mathrm{s}}{\mu_\mathrm{solvent}}\sigma_\mathrm{w}(t). 	
\end{equation}
The slip velocity is incorporated into equation (2), and the value of $l_\mathrm{s}$ is determined by fitting the slope in the stress-velocity plot shown in Fig. \ref{fig:Fig2}(c). The slip gradually develops even before solidification when $h<h_\mathrm{p}$, but the slip's effect on the rising speed is negligible as $\sigma _ \mathrm{Y} \ll \sigma _ \mathrm{w} (t)$. After the solidification, the slip itself takes on the role of the apparent speed as YSF is not flowing, considering the contribution from equation (2) is zero with $\sigma _ \mathrm{w} (t)<\sigma _ \mathrm{Y}$. Due to the discontinuity of equation (2) between regimes before and after $h_\mathrm{p}$, the numerical calculation explicitly connects the two regimes. The final solution of the early regime, which includes $h, \dot{h},$ and $\sigma_\mathrm{w} $, serves as the initial condition after $h_\mathrm{p}$.

\begin{figure}[t]
\includegraphics[width=1\linewidth]{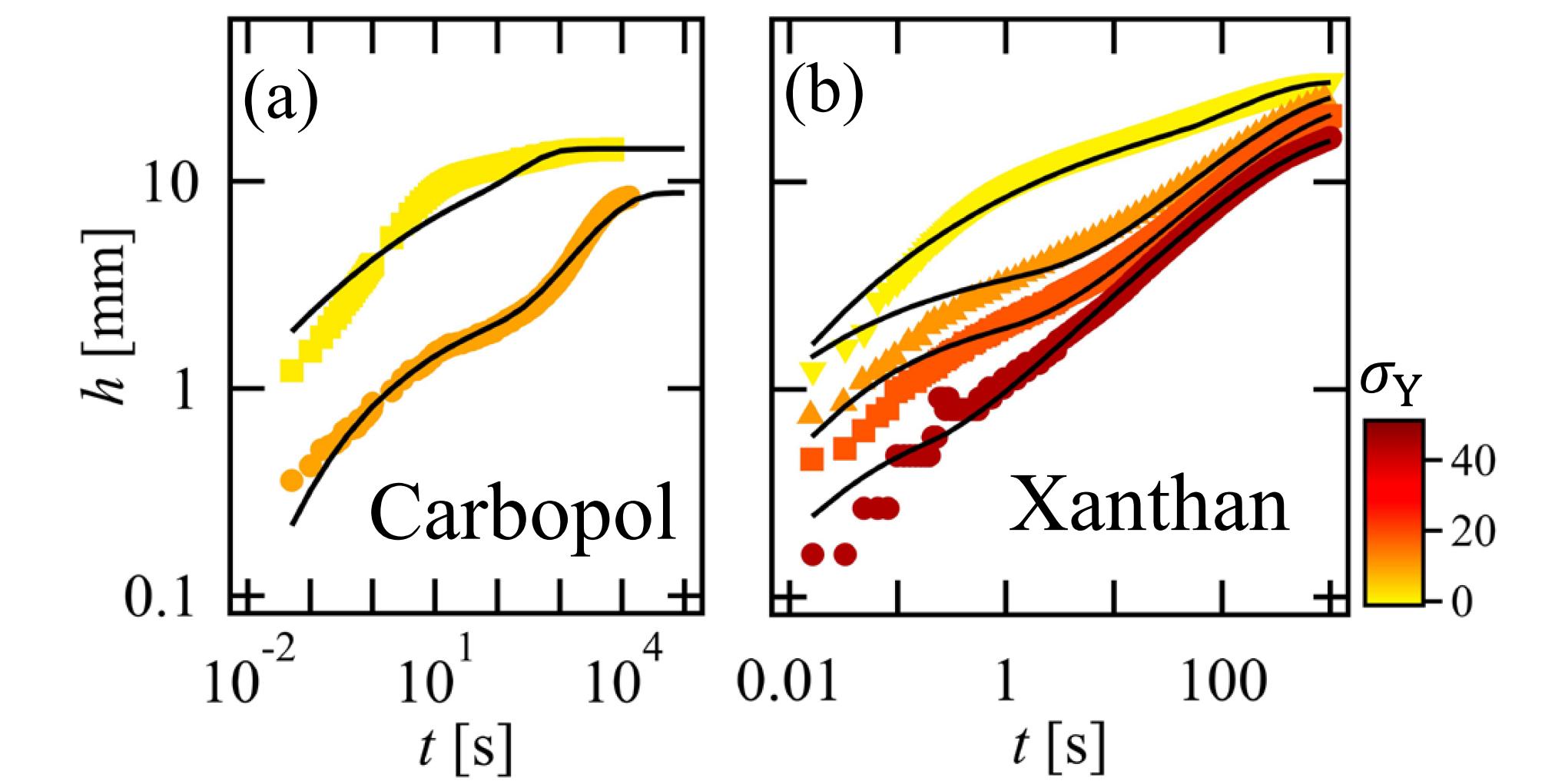}
\caption{\label{fig:Fig3}Proposed model prediction with (a) Carbopol and (b) Xanthan solutions. Solid lines are numerical solutions of proposed model (equation (1) – (3)). From top to bottom, concentrations are 0.08 and 0.12 wt. $\%$ in (a), while 0.3, 0.7, 1.0, and 1.3 wt. $\%$ in (b). The marker color is presented in colormap of yield stress. }
\end{figure}

In Fig. \ref{fig:Fig3}, the solid lines represent the capillary rise prediction of equations (1)-(3) for different YSFs. The experimental data is plotted using color markers to indicate the strength of the yield stress. Using the measured rheological properties and the estimated surface tension, $l_\mathrm{s}$ is calculated from the slope of a characteristic plot displayed as Fig. \ref{fig:Fig2}(c). A detailed fitting result is given in \cite{SI}. Surprisingly, model curves are capable of effectively capturing the diverse dynamics of capillary rise. The transition from solidification to gliding flow is accurately described as smooth, as supported by experimental results. The solidifying plateau does not completely flatten ($\dot{h}=0$) in the middle of the curve because the slip is already in action even before solidification.

To better reflect the initial trajectory of 1.3 and 1.0 wt. $\%$ Xanthan solutions, the yield stress was empirically fitted from the rise curve. This is due to the ambiguity in the yield stress determination for Xanthan \cite{Tang2018,Moller2009,Barnes1985}. The residual discrepancy can be explained with shear banding as proven in \cite{SI}. By reducing shear banding effect with salt\cite{Tang2018}, experimental data follows the proposed model very well. To be within the scope of this work, accurate prediction of shear banding or thixotropic property \cite{Moller2009,DeSouzaMendes2011,Divoux2016} is not included here. Although Carbopol 0.08 wt. $\%$ sample’s yield stress is also fitted, a plateau curve appears and could be interpreted with our proposed model, when the surface is altered into a less slippery condition through surface silanization\cite{SI}. Issues related to thicker YSFs have been addressed in \cite{SI}.

\begin{figure}[t]
\includegraphics[width=1\linewidth]{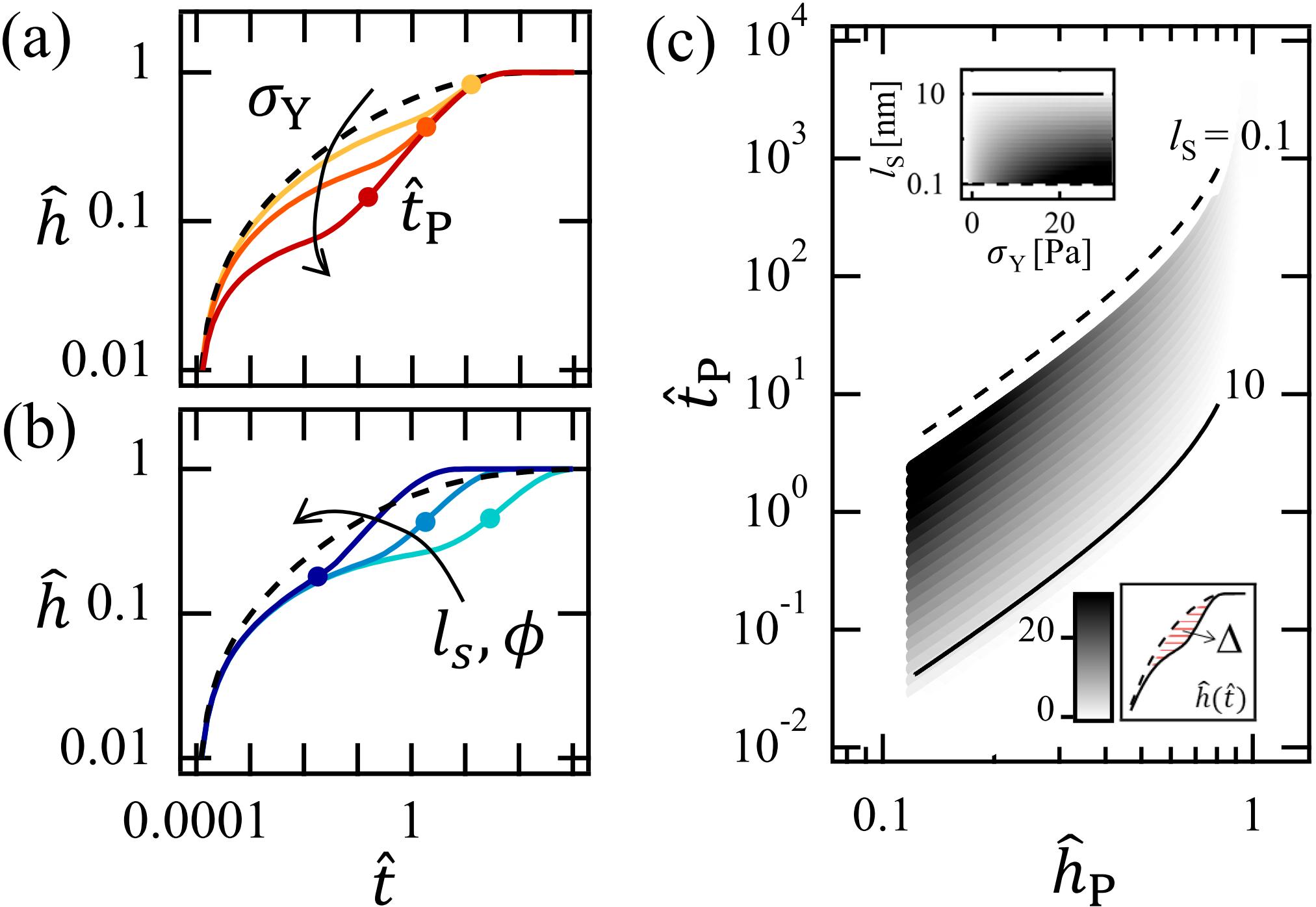}
\caption{\label{fig:Fig4}(a) Normalized capillary rise curves varying $\sigma_\mathrm{Y}$. From the bottom, $\sigma_\mathrm{Y}$ = 40, 10, and 2 Pa. (b) Normalized capillary rise curves varying $l_\mathrm{s}$, the predominance of slip. From the bottom, $l_\mathrm{s}$ ($\phi$) = 0.01 (0.006), 0.2 (0.1), and 16 (9) nm. Dashed lines in (a) and (b) are shear thinning curves without slip. Circle marker on line denotes $\hat{t_\mathrm{p}}$. (c) Contour plot of $\Delta$, the intensity of the plateau, shown against normalized plateau time and height. The solid line shows $l_\mathrm{s}$ = 10 nm and the dashed line is 0.1 nm. The lower inset is the graphical definition of $\Delta$. The upper inset shows same contour over $l_\mathrm{s}$ and $\sigma_\mathrm{Y}$. For all graphs, conditions other than $l_\mathrm{s}$ and $\sigma_\mathrm{Y}$ are constraint as Carbopol 0.12 wt. $\%$.\cite{SI}}
\end{figure}

The plateau shows diverse intensities and shapes both in our experiments and theoretical model. It can be expected that a rise of fluids with larger yield stress results in a stronger plateau. The plateau at 1.3 wt. $\%$ Xanthan solution, however, is less pronounced and nearly unnoticeable as compared to the lower concentrations. It can be inferred that the strength of a plateau might be determined not solely by the yield stress, but by other variables such as slip and consistency parameter ($K$).

Based on the proposed model, the various intensities and shapes of the plateau are studied. Capillary rise curves were obtained in Fig. \ref{fig:Fig4}(a) and (b) by changing only the rheology or slip, respectively, in order to distinguish the effects from other factors. All axes are normalized, where $\hat{h}$ is the rise height divided by the final height ($h_\mathrm{E}$) and $\hat{t}$ is the time divided by the characteristic time scale of capillary rise ($t_\mathrm{v}$, which is defined without yield stress) \cite{Kim2020}.

In Fig. \ref{fig:Fig4}(a), an increase in the yield stress gives rise to more rapid solidification to occur. It also slows down the initial rise because the fluid's yield stress raises the apparent viscosity. Fig. \ref{fig:Fig4}(b) depicts the variation of slip strength while assuming a constant yield stress. In order to assess the extent of slip at the plateau, the velocity ratio ($\phi$) is obtained by dividing slip contribution ($V_\mathrm{s}$) by the non-slip contribution ($V_\mathrm{ns}=\dot{h}-V_\mathrm{s}$). Around the plateau ($\sigma_\mathrm{w}=2\sigma_\mathrm{Y}$),
\begin{equation}
\phi={\frac{V_{\mathrm{s},\epsilon=0.5}}{V_{\mathrm{ns},\epsilon=0.5}}}=\frac{8g\left(n\right)l_\mathrm{s}}{R\mu_\mathrm{solvent} } {\left(\frac{K}{{\sigma_\mathrm{Y}}^{1-n}} \right)}^{1/n},  	
\end{equation}
where $g(n)=\frac{(n+1)(2n+1)(3n+1)}{n(7n^2+8n+2)}$. The two graphs (Figs. \ref{fig:Fig4}(a) and (b)) are completely different, suggesting that each has a unique impact on the capillary rise dynamics. The position of the plateau and the initial rate of rise are mainly determined by the yield stress, whereas the persistence of solidification and the rate of later rise are affected by slip.

 The time scale for the ending of plateau ($t_\mathrm{p}$) is derived via dividing plateau height by slip velocity, as 
\begin{equation}
t_\mathrm{p}=\frac{h_\mathrm{p}}{V_{s,\sigma_\mathrm{w}=\sigma_\mathrm{Y}}} =\frac{h_\mathrm{E} \mu_\mathrm{solvent}}{l_\mathrm{s}\sigma_\mathrm{Y}\left(1+\frac{2\sigma_\mathrm{Y}}{\rho gR}\right)}.
\end{equation}
$\hat{t_\mathrm{p}}$ is $t_\mathrm{p}$ divided by $t_\mathrm{v}$ and is circled on the solid line in Figs. \ref{fig:Fig4}(a) and (b). Solidification finishes faster for higher yield stress and more slippery cases. It is interesting to note that when the $l_\mathrm{s}$ reduces down below a single atomic size, $t_\mathrm{p}$ prolongs longer than $10^3$ s. According to \cite{Zhang2017}, this time frame is sufficient for the evaporation-induced residual  stress to build at the edge. The evaporation at the edge slows down the capillary imbibition speed because it is focused mainly on the wetting friction of meniscus \cite{Zhao2019}. In Young's spreading of YSFs on rough surfaces, the similar cessation occurs if the plateau endures for a sufficient duration, allowing the meniscal edge to undergo complete drying \cite{Jalaal2021}. Unless evaporation is controlled, YSFs could be easily stopped, leading to a false conclusion.

For a general understanding of the plateau intensity, we evaluated the flexion of plateau ($\Delta$) as
\begin{equation}
\Delta = \int_{0}^{\infty}{\tilde{h_0}\left(\tilde{t}\right)-\tilde{h}\left(\tilde{t}\right)}\,d\tilde{t} ,    
\end{equation}
where $\tilde{t}$ and $\tilde{h}$ are logarithmic $\hat{t}$ and $\hat{h}$; $\tilde{h_0}$ is normalized logarithmic rise height with $\sigma_\mathrm{Y}=0$. As shown in the bottom inset $h(t)$ curve of Fig. \ref{fig:Fig4}(c), $\Delta$ is the integration of the difference in height between curves $\tilde{h_0}$ and $\tilde{h}$. The intensity is calculated varying $\sigma_\mathrm{Y}$ and $l_\mathrm{s}$ as shown on the plane of normalized plateau time and height. According to the presented phase diagram, it can be observed that the plateau strength is positively correlated with the magnitude of the yield stress, while exhibiting an inverse relationship with the degree of slip. Even if the solidification occurs early (low $\hat{h_\mathrm{p}}$), their flow can closely resemble dynamics of shear-thinning fluids when they experience a large slip effect (low $\hat{t_\mathrm{p}}$), as well as in high $\hat{h_\mathrm{p}}$ cases. The enhancement of slip can be obtained in cases of the increasing thickness of the slip layer (e.g., the enhancing repulsive forces between the YSF element and the smooth wall).

We revisited the capillary imbibition problem with yield stress fluids. YSFs inevitably solidify under such wetting or spreading flow where the speed continuously slows down. The liquid-solid transition appears as a plateau at the rise curve, at glance, but its re-rise occurs by a slip between YSF and inner wall surface. By assuming a depleted slip layer, the entire capillary rise curve could be predicted with a simple rheology model. The MATLAB code utilized for numerical calculations is readily accessible \cite{SI}. The suggested model including DCA theory can provide crucial information about how YSFs  flow inside channels at lower stresses, which is difficult to accurately measure, typically requiring a complicated force sensor or a micro-PIV device \cite{Jorgensen2017,Jorgensen2015,Graziano2021,Pemeja2019}. Our capillary imbibition model with experiments can be helpful to obtain parameters such as surface tension, slip length, and dynamic wetting coefficients, required for wetting, spreading, simple pipe flow, and any other slippery flow. Since predicting penetration rate and fluid uptake in porous media shares the same phenomena with capillary rise, our system can also provide clues for modeling YSF flow in porous media.

\begin{acknowledgments}
We are grateful to Sungsoo Han and Korea Institute of Geoscience and Mineral Resources (KIGAM) for helpful discussions and Kyu Han Kim and Kyoungmun Lee for their comments on the manuscripts. This research was supported by Basic Science Research Program through the National Research Foundation of Korea (NRF) (NRF-2021R1A6A3A13046176, NRF-2021R1A2C2009859) and Korea Evaluation Institute of Industrial Technology (20014762).
\end{acknowledgments}

%

\end{document}